\documentclass[pdflatex,sn-nature]{sn-jnl}

\usepackage{xr}
\usepackage{graphicx}%
\usepackage{multirow}%
\usepackage{amsmath,amssymb,amsfonts}%
\usepackage{amsthm}%
\usepackage{mathrsfs}%
\usepackage[title]{appendix}%
\usepackage{xcolor}%
\usepackage{textcomp}%
\usepackage{manyfoot}%
\usepackage{booktabs}%
\usepackage{algorithm}%
\usepackage{algorithmicx}%
\usepackage{algpseudocode}%
\usepackage{listings}%
\usepackage{gensymb}

\newcommand{\vstar}[0]{\(V^*\)}
\newcommand{\hstar}[0]{\(H^*\)}
\newcommand{\vhstar}[0]{\(\bigl(V^*,H^*\bigr)\)}

\begin{document}

\title{Fabrication-Directed Entanglement for Designing Chiral and Anisotropic Metamaterial Foams}


\author[1]{\fnm{Daniel} \sur{Revier}}\email{drevier@uw.edu}

\author*[2]{\fnm{Jeffrey} \sur{Lipton}}\email{j.lipton@northeastern.edu}


\affil[1]{\orgdiv{Paul G. Allen School of Computer Science and Engineering}, \orgname{University of Washington}, \orgaddress{\street{1410 NE Campus Parkway}, \city{Seattle}, \postcode{98195}, \state{WA}, \country{USA}}}

\affil[2]{\orgdiv{Department of Mechanical and Industrial Engineering}, \orgname{Northeastern University}, \orgaddress{\street{390 Huntington Ave.}, \city{Boston}, \postcode{02115}, \state{MA}, \country{USA}}}



\abstract{Entangled networks are fundamental in various systems, from biological structures to engineered materials. 
Current techniques for programming entanglement often rely on intricate chemistry or result in statistically homogeneous networks, limiting the ability to create spatially patterned structures with precisely engineered functions. 
Thus, a key challenge remains in developing approaches to program complex mechanical behaviors, such as anisotropy and chirality, within monolithic entangled structures. 
This work introduces Fabrication-Directed Entanglement (FDE), a methodology integrating viscous thread printing (VTP) and topology optimization (TO) to program the entanglement of a single homogeneous filament.
By spatially adjusting VTP parameters (deposition height, speed), we control local coiling density, creating quasi-two-phase (dense/sparse) regions within a monolithic entangled foam. 
Topology optimization guides the placement of these regions to achieve target macroscopic mechanical properties. 
Here we show foam-like mechanical metamaterials with tunable compliance, rigidity, and chirality. 
Experimental testing and simulation confirm that FDE expands the achievable material property space compared to homogeneous VTP foams, enabling properties like tunable directional stiffness, Poisson's ratios from $\nu\approx0.06~\text{to}~0.56$, and novel significant normal-shear coupling ($\eta_{212}\approx0.72$) from a single base material. 
This approach offers a viable new pathway for designing complex, functional entangled foam structures with tailored mechanical behaviors.}

\keywords{Metamaterials; Chirality; Foams; Topology Optimization; Viscous Thread Printing (VTP)}



\maketitle

\section{Introduction}\label{intro}
Physically entangled networks underpin a variety of essential functions across length-scales, from protein folding and polymer assemblies to larger biological and engineering systems \cite{Salicari2023-bs, Dhand2024-rl, Lu2008-ei, Day2024-tu}. 
At the microscale, the self-interaction is typically encoded directly by the sequence of its chemical composition~\cite{Sulkowska2020-qk, Dabrowski-Tumanski2017-yl, Christian2016-ch}
In contrast, methods that produce macroscopic entangled materials such as entangled metallic wire materials~\cite{Zhang2013-wy, Tan2009-ce, Tan2012-mu, Liu2010-hf, Gadot2015-el, Rodney2015-jq} and non‐woven or electrospun fiber mats \cite{Kuts2024-pn, Kumbar2008-tn, Xue2019-pd, Li2004-nu} lack the fine‐grained spatial controls needed to prescribe local entanglement patterns, limiting the ability to tailor macroscopic mechanical response.
Thus, current techniques either rely on intricate, microscale chemistry or yield statistically homogeneous macroscopic networks---leaving an important gap in our ability to create spatially patterned structures that achieve precisely engineered functions.

Here we introduce fabrication-directed entanglement (FDE), a new approach leveraging computational design (Topology Optimization, TO~\cite{Bendsoe1988-hx}) and additive manufacturing (Viscous Thread Printing, VTP~\cite{Lipton2016-ou, Yuk2018-br}) to create spatially patterned entanglement within macroscopic structures, targeting prescribed elastic behaviors.
FDE uniquely enables the programming of complex mechanical responses like targeted anisotropy and chirality \emph{within} a continuous, single-material entangled foam structure---a capability previously unrealized in this material class.
While the underlying VTP process involves certain trade-offs, notably resolution limits ($\sim1~\text{mm}$) and moderate achievable stiffness contrast ($\sim30{:}1$) compared to some other additive multi-material manufacturing techniques, this integrated approach allows TO to strategically guide VTP coiling density, placing regions of dense or sparse entanglement to achieve the targeted macroscopic response.


In contrast to sequence-based chemical encoding or statistically homogeneous fabrication methods, FDE enables direct control over the local entangled network structure during VTP deposition, offering spatial modulation of density and connectivity from a single filament. 
This programmability is key: it allows tailoring the macroscopic mechanical response in ways inaccessible through simple material mixing or uniform VTP structures, demonstrating the value of programmed heterogeneity even within the inherent constraints of the VTP system. 
Furthermore, the resulting monolithic foam structure provides a direct physical analogue to TO's continuum assumption, potentially mitigating connectivity challenges sometimes faced in discrete solid-void realizations. 
The practical outcome is therefore a broadened design space for monolithic entangled foams, directly evidenced later in this work by achieving targeted directional stiffness and significant programmed normal-shear coupling (\autoref{sec:anisotropy}, \ref{sec:normal-shear})---complex behaviors arising from the FDE-patterned network structure.


In this work, we systematically explore how higher‐order, spatially controlled entanglement patterns affect the mechanical response of 2D foam materials. 
We show that FDE‐based structures surpass simple mixtures of homogeneous materials in terms of accessible property space, establishing FDE as a method for programming mechanical properties through controlled deposition. 
Specifically, we:
\begin{enumerate}
    \item Introduce and implement the FDE method, using computational guidance (via TO) to inform VTP deposition.
    \item Experimentally validate the FDE workflow by fabricating TO designs and demonstrating targeted mechanical behaviors like anisotropy and chirality.
    \item Show, through simulation and experimental examples, an expanded property space compared to uniform VTP foams, including programmable directional stiffness and normal-shear coupling.
\end{enumerate}

In what follows, we first provide background on VTP and TO, highlighting the limitations that motivate our approach.
We then detail our FDE methodology and subsequently present experimental and computational results which demonstrate FDE's ability to achieve selective compliance, stiffness, and chirality, expanding the available property space beyond homogeneous VTP foams. 
Finally, we discuss the implications, limitations, and broader potential of this approach.

\begin{figure}[tbp]
    \centering
    \includegraphics[width=0.9\linewidth]{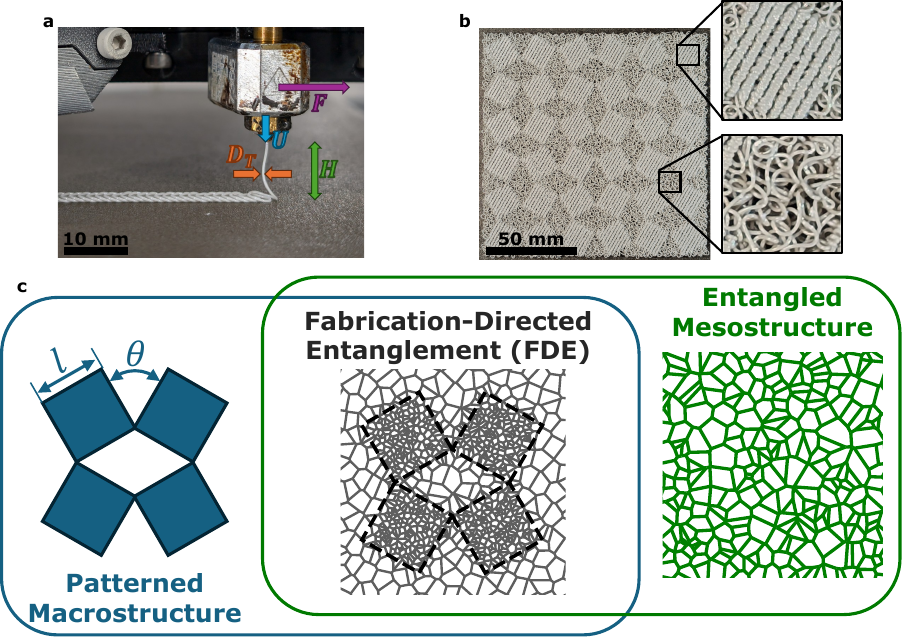}
    \caption{a) The VTP process depositing a line of stacked coils with key parameters highlighted: $F$ (translational speed), $U$ (extrusion speed), $H$ (print height), and $D_T$ (thread diameter). b) A ``rotating squares’’ metamaterial produced with FDE by changing the coiling parameters. Zoomed insets show the different densities and characteristics of dense and sparse coiling. c) The FDE process takes an explicitly defined macrostructure and combines it with an implicit stochastic microstructure (schematically represented here by a Voronoi diagram), to produce variable-density entangled structures.}
    \label{fig:intro}
\end{figure}

\section{Background}
FDE builds upon established methods of VTP and TO via inverse homogenization (IH). 
This section reviews the key developments and limitations in these precursor fields to contextualize the FDE approach presented in this work.

\subsection{Viscous Thread Printing}
VTP is a three-dimensional extension of the viscous thread instability (VTI) phenomenon. 
VTI is  likened to drizzling honey onto toast; when honey is drizzled from a short height it spontaneously coils and loops on the toast rather than settling in a neat line. 
VTP applies this same phenomenon to molten, semi-viscous filaments from a 3D printer, but under precise control of nozzle motion and extrusion parameters. 
By systematically controlling the coiling, a single continuous strand can self-interact to form entangled, foam architectures with programmable rigidity and compliance \cite{Emery2021-xm, Emery2024-bb}.

We parameterize the coiling via two dimensionless parameters---height \hstar{} and speed \vstar{}---which broadly determine coil radius and coil period, respectively \cite{Yuk2018-br, Brun2015-qm}. 
By adjusting these parameters, the filament can be made to form dense, tightly wound loops or looser, more open coils---thereby controlling stiffness.
In practice, these parameters are defined as
\[
H^* \;=\; \frac{H}{D_T}
\quad\text{and}\quad
V^* \;=\; \frac{F}{U},
\]
where \(H\) is the nozzle height above the substrate, \(D_T\) is the extruded filament diameter, \(F\) is the nozzle translation speed, and \(U\) is the filament extrusion speed at the nozzle exit (\autoref{fig:intro}a).
With a standard extrusion-based 3D printer, we control the coiling by varying \vstar{} and \hstar{} via the print height $H$ and nozzle speed $F$, while keeping the nozzle thread $D_T$ and extrusion speed $U$ fixed.
This enables us to produce a wide range of characteristic coiling with minimal control parameters. 

Previous VTP work has predominantly focused on statistically-homogeneous structures (i.e., constant \vhstar{} in a print domain) to investigate the effects of \vhstar{} on the overall stiffness of VTP foams~\cite{Lipton2016-ou, Emery2021-xm}.
More recently, there has been a preliminary investigation into spatially patterning stiffness using gradients~\cite{Emery2023-jx} in addition to developing a model to predict modulus and effective layer height given a \vhstar{} pair~\cite{Emery2024-bb}.
Surprisingly, this latest work found a space of \vhstar{} pairs that are compatible in a single print process and thus can be used to spatially program stiffness with stiffness ratios on the order of \(10^2{:}1\).
This illuminates a design space that can be exploited to create entangled, quasi-two-phase metamaterials by spatially patterning different stiffnesses; however, does not inform how the space should be patterned to implement desired mechanical responses.

\subsection{Inverse Homogenization}
IH is a computational technique specifically developed for designing material microstructures. 
It applies TO principles, combined with periodic homogenization, to find the optimal layout within a unit cell that yields desired effective macroscopic properties~\cite{Sigmund1994-qf, Andreassen2014-en, Andreassen2014-tc, Bendsoe1988-hx}.
This powerful technique is widely used in metamaterial design across various domains, including mechanical, acoustic, electromagnetic, and optical applications~\cite{El-Kahlout2011-kw, Diaz2010-oi, Lu2013-lr, Jensen2011-ra}.

The core principle involves optimizing an objective function based on the homogenized material properties, typically represented by the constitutive equation of the underlying physics. 
For instance, in mechanical metamaterials, IH is typically used to optimize functions based on the microstructure’s symmetric homogenized elasticity matrix $C$, given by the linear elastic relationship $\sigma=C\varepsilon$.

This approach has proven effective for discovering ``extremal materials,'' whose behavior reaches the theoretical bounds of linear elasticity and is ultimately governed by the eigenvalues and eigenvectors of $C$ \cite{Milton1995-rd, Sigmund1994-qf, Andreassen2014-tc}. 
These materials are classified by their number of compliant directions (i.e., small eigenvalues): materials with exactly one compliant direction are termed ``unimodes,'' those with two are termed ``bimodes,'' and so on up to ``trimodes'' for 2D and ``hexamodes'' for 3D linear elastic materials. 
In 2D linear elastic materials a nullmode is trivially rigid and a trimode is trivially compliant, hence this work focuses on unimode and bimode designs.

Prior studies often achieved such designs by targeting specific bulk material properties to reach theoretical bounds---for example, minimizing Poisson's ratio (yielding isotropic unimodes) or maximizing the bulk-to-shear modulus ratio (yielding isotropic pentamodes)~\cite{Andreassen2014-tc, Kadic2012-qy}. 
Focusing only on scalar properties, however, under-specifies the desired mechanics and often requires additional constraints (e.g., symmetry, minimum stiffness) to yield meaningful designs. 
While the full elasticity matrix $C$ has been considered in prior work, its use has often focused on achieving specific orthotropic or isotropic symmetries~\cite{Sigmund1994-qf, Andreassen2014-tc}, rather than targeting directed anisotropy or complex coupling effects.
Thus, designing microstructures that specifically target arbitrary $C$ matrices remains unexplored at this time.

Additionally, prior work in IH for mechanical metamaterials often assumes high stiffness contrast between constituent material phases, with ratios spanning \(10^3{:}1\) to \(10^9{:}1\) commonly used to achieve extremal properties~\cite{Bendsoe2013-fo}. 
The contrast pushes optimization towards binary (solid-void) designs, as these avoid the significant fabrication complexities associated with intermediate stiffness values or multiple distinct phases using traditional methods. 
However, while simplifying the phase requirements, this focus on high-contrast, binary designs introduces manufacturability challenges common to density-based optimization outputs \cite{Andreassen2014-tc, Wang2011-gy, Hammond2021-ir}. 
Optimized layouts can exhibit artifacts like indistinct boundaries (``gray areas''), checkerboard patterns, or tenuous connections, requiring regularization or constraints to ensure robust microstructures. 
Thus, it remains an open question how a simpler, continuous fabrication method might overcome these combined limitations.

\section{Fabricated-Directed Entanglement}
Our FDE methodology, illustrated in~\autoref{fig:intro}, integrates VTP and IH to impose higher-order patterns onto the inherently stochastic VTP coiling process. 
By spatially varying VTP parameters according to an IH-derived design, we create quasi-two-phase entangled structures with programmed local density and tailored macroscopic properties. 

\subsection{Viscous Thread Printing (VTP) Enables Control of Local Entanglement}
As highlighted previously, we can control characteristic entanglement (dense vs. sparse) via VTP parameters \hstar{} and \vstar{}, controlled by nozzle height $H$ and speed $F$ when filament diameter $D_T$ and extrusion speed $U$ are fixed, thereby tuning local stiffness independently of the global toolpath.
We selected to compatible \vhstar{} pairs (\autoref{method:mat_and_fab}) yielding dense and spare ``phases'' with an approximate measured stiffness ratio of $\sim30{:}1$. 
Additionally, the selection of the print parameters resulted in a minimum feature width $\sim1.2$ mm.

By spatially varying these parameters during the build, dense or sparse regions can be patterned on demand; however, the inherent coiling dynamics of VTP involve a trade-off.
While gaining complex filament entanglement as an emergent property of the deposition dynamics, VTP cannot resolve features (e.g., hinges) smaller than the characteristic filament coil size.
Furthermore, the limited stiffness ratio of the dense and sparse coils ultimately restricts the possible range of achievable material properties.
Despite these limitations, with a single, homogeneous thread, this process yields a quasi-two-phase structure that expands the possible range of mechanistic behaviors of monolithic entangled foam metamaterials---particularly anisotropy and chirality.

\subsection{Inverse Homogenization Enables Higher-Order Patterning of Local Entanglement}
\label{sec:IH}
To strategically pattern the dense and sparse entangled regions created by VTP, we employ an IH framework following conventional density-based TO implementations in mechanical metamaterials \cite{Bendsoe2013-fo} with two key differences: 1) we interpret TO low and high density regions as sparse and dense entanglements and 2) we optimize for a desired eigenstructure (i.e., principal directions and stiffnesses) instead of specific material properties.

First, the optimized density field $\rho$ directly informs the VTP process, where $\rho=0$ corresponds to the sparse, compliant \vhstar{} pair, and $\rho=1$ corresponds to the dense, rigid \vhstar{} pair (\autoref{fig:intro}b). 
Whereas conventional TO often uses a solid-void material set for the density field, with stiffness contrasts such as $10^9{:}1$, our FDE approach operates with a quasi-two-phase entanglement system with stiffness $\sim30{:}1$.
While this stiffness ratio is substantially lower than conventional TO work, restricting the available property space, it is nevertheless comparable to those found in traditional composites~\cite{Jones2018-yl}, indicating that an engineered response is achievable.
Furthermore, since FDE translates the density map into variations along a single, continuous filament path via VTP, the resulting structure is inherently continuous and well-connected---directly realizing the domain continuity in the TO FEM formulation---and is thus free from the disconnected features or ill-defined interfaces that can plague conventional TO.

Second, to target complex material behaviors we develop a generalized framework based on principal deformation modes (eigenstructure) ($V,~\Lambda$), where $T^*=V\Lambda V^T$.
This approach, inspired by extremal materials theory~\cite{Milton1995-rd}, allows designers to directly specify desired principal directions ($V$) and their relative stiffnesses ($\Lambda$) rather than deriving them from bulk properties. 
This direct specification greatly facilitates the exploration of diverse anisotropic and coupled behaviors; for instance, the chiral designs herein emerged unexpectedly when targeting specific normal-shear coupled strains using this method. 

Here $V\in\mathbb{R}^{3\times3}$ where $V^TV=I$ contains the desired principal directions (a design input) while the matrix $\Lambda=\text{diag}(a, 1, 1)$ for $a \ll 1$ defines the target eigenvalues. 
Throughout this work, the eigenvector strains of $V$ are expressed using as $v_i=[\varepsilon_{11}, \varepsilon_{22}, \gamma_{12}]^T$ and the specific library of $V$ matrices used in this work is provided in the SI (Table S2). 
Thus by design, $T^*$ possesses one relatively small eigenvalue corresponding to the first column vector of $V$, which we term the ``leading direction,'' $\mathbf{v}_{lead}$, with remaining columns corresponding to larger eigenvalues and span the orthonormal ``supporting directions.'' 
Importantly, the eigenstructure also implicitly encodes material symmetry, biasing the optimizer for certain topological symmetries, but not explicitly enforcing them.
 
The core optimization then seeks to match the eigenstructure of the homogenized material properties $M$ ($M=C$ for unimodes and $M=S=C^{-1}$ for bimodes) to this target $T^*$. 
The homogenized stiffness matrix $C$ is used directly for unimodes, as targeting the single compliant direction of $v_{lead}$ fits the definition of a unimode material and simultaneously drives rigidity in the supporting directions. 
For bimodes requiring two compliant directions, we instead use the homogenized compliance matrix $S$. 
This sets $v_{lead}$ to be a direction of rigidity, not compliance, and therefore drives the two supporting directions to now be directions of compliance.
Specifically, we solve:
\[
\begin{aligned} \min_{\rho} \quad & \lVert \hat{M} - T^* \rVert_F^2 \ \\
\text{s.t.} \quad & K(\rho)u = f \\
\quad & g_n(\rho) \le 0
\end{aligned}
\]
Here, $K(\rho)u=f$ is the static equilibrium constraint, $g_n(\rho)$ is any number of additional desired constraints (e.g., volume fraction, explicit symmetry), and $\hat{M} = M / \lVert M \rVert_2$ is the normalized homogenized matrix, which allows the optimizer to focus on aligning the eigenstructure irrespective of the overall material distribution and stiffness.

FDE combines VTP and IH to solve the joint problems of local entanglement and its spatial distribution.
This results in a system that achieves directed mechanical performance by encoding characteristic entanglement as an optimized density field using targets defined by extremal material theory.
While the moderate stiffness contrast achievable with VTP precludes reaching true extremal bounds, FDE employs topology optimization guided by the principles from extremal theory (e.g., targeting specific eigenstructures) to effectively utilize the available contrast for designing foams with programmed anisotropy and chirality.

\section{Results \& Discussion}
\begin{figure}[htbp]
    \centering
    \includegraphics[width=\linewidth]{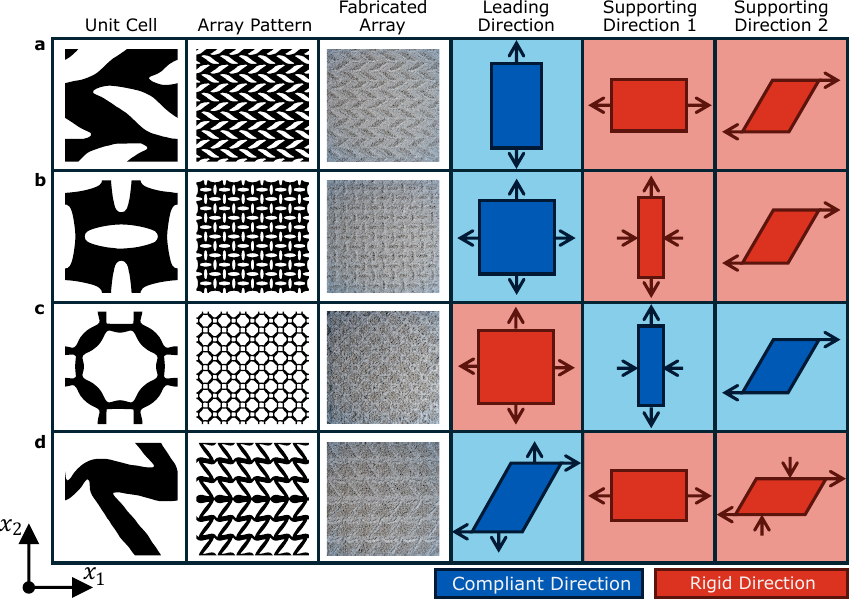}
    \caption{
      Topologies generated by optimization (pre-calibration, shown to illustrate the target fine features before accounting for fabrication limits discussed in~\autoref{method:calibration}) for materials targeting different modes of compliance (blue) and rigidity (red).
      a) Unimodal, compliant to vertical strain, rigid otherwise.
      b) Unimodal, compliant to bulk strain, rigid to shear.
      c) Bimodal, rigid to bulk strain, complaint to shear.
      d) Unimodal, compliant to positive vertical-shear coupling. This topology was mirrored vertically during fabrication to manage testing boundary conditions.
    }
    \label{fig:array_modes}
\end{figure}

We empirically demonstrate our FDE method by fabricating and measuring the properties of four optimized samples, shown in~\autoref{fig:array_modes}. 
These designs represent different numbers of compliant or rigid modes and leading directions \(\mathbf{v}_{lead}\).
All samples were fabricated and tested under uniaxial compression (see Methods).
Given their relatively thin geometry (aspect ratio approx. 5.4:1 based on the $150\times150\times28~\text{mm}^3$ dimensions), the samples predictably exhibited global buckling under load at approximately $5\%$ strain.
Consequently, the effective elastic properties (Young's moduli $E_1, E_2$; Poisson's ratios $\nu_{ji} = -\epsilon_{ii}/\epsilon_{jj}$; and normal-shear coupling coefficients $\eta_{kij} = \gamma_{ij}/\epsilon_{kk}$) were extracted using data solely from the initial $0-2.5\%$ strain range, prior to the onset of this instability.
This strain range provided a clear linear elastic regime across all samples, suitable for determining effective properties via linear regression fits to the measured data.
It should be noted, however, while this approach allows for characterizing the material's intrinsic linear response, the early buckling onset suggests that the operational strain range of these specific large-aspect-ratio structures is likely limited in unconfined compression for applications requiring larger strains; achieving stability at higher strains could require geometric modifications or the application of external constraints.

\autoref{tab:meas_sim_data} validates the FDE process by comparing the measured effective properties against simulation results that were calibrated post-fabrication  by adjusting the TO projection dilation factor $\chi$ to account for fabrication minimum feature size (see Methods~\autoref{method:calibration} for full details).
This calibration accounts for the scale-dependent effect of our VTP resolution limits ($\sim1.2~\text{mm}$ coil diameter), which cause effective thickening of fine features in the fabricated samples compared to the idealized topology.
Overall, the table shows reasonable agreement between the measured and simulated properties across the different designs.

\begin{table}[htbp]
\centering
\caption{Measured versus simulated effective properties for the four design examples depicted in \autoref{fig:array_modes}. Values are reported as Measured(Simulated). $E$ values are normalized to the dense phase modulus ($\sim 3.4~\text{MPa}$). Simulated values are for calibrated topologies (see~\autoref{fig:mat_props}).}
\label{tab:meas_sim_data}
\setlength{\tabcolsep}{4pt}
\small
\begin{tabular}{@{}lcccccc@{}}
\toprule
Family & $E_1$ & $E_2$ & $\nu_{12}$ & $\nu_{21}$ & $\eta_{112}$ & $\eta_{212}$ \\ 
\midrule
Unimode, Vertical & $0.54(0.48)$ & $0.29(0.29)$ & $0.24(0.38)$ & $0.15(0.22)$ & $0.08(-0.01)$ & $0.00(0.00)$ \\
Unimode, Bulk & $0.61(0.58)$ & $0.47(0.46)$ & $0.06(0.18)$ & $0.06(0.14)$ & $0.03(0.00)$ & $0.03(0.00)$ \\
Bimode, Bulk & $0.23(0.24)$ & $0.22(0.24)$ & $0.56(0.57)$ & $0.52(0.57)$ & $0.02(0.00)$ & $0.02(0.00)$ \\
Unimode, Normal-Shear & $0.37(0.41)$ & $0.21(0.19)$ & $0.30(0.31)$ & $0.14(0.14)$ & $0.03(0.14)$ & $0.72(0.67)$ \\
\bottomrule
\end{tabular}
\end{table}

\subsection{Targeted Anisotropy and Bulk/Shear Response}\label{sec:anisotropy}
The first example (\autoref{fig:array_modes}a) was optimized to be unimodal: compliant in the vertical direction while remaining rigid under horizontal and shear strains. 
Structurally, it features thick, diagonally oriented members connected by thin horizontal hinges. 
Testing yielded an elastic modulus ratio \(E_1/E_2\approx 1.87\), $\nu_{12} \approx 0.28$, \(\nu_{21}\approx 0.15\), and negligible normal-shear coupling (\autoref{tab:meas_sim_data}), demonstrating clear orthogonal anisotropy, consistent with the design goal. 
The ability to achieve such targeted anisotropy, isolating deformation primarily to one direction, highlights FDE's capacity to move beyond the typically isotropic behavior of statistically homogeneous foams and is valuable in scenarios requiring precision motion control or minimal lateral bulging (e.g., mounts, seals).

The second test piece (\autoref{fig:array_modes}b) targeted unimodal compliance under bulk strain while remaining rigid to shear, yielding a topology resembling ``rotating squares''\cite{Grima2000-ln}. 
Experimental testing confirmed the low Poisson's ratio goal $(\nu_{12} \approx \nu_{21} \approx 0.06)$, although it also revealed moderate elastic anisotropy $(E_1/E_2 \approx 1.29)$. 
Nevertheless, this resulting anisotropy underscores that the $T^*$-guided optimization allows flexibility; while biasing towards certain symmetries, it doesn't strictly enforce them, permitting anisotropic solutions even when more symmetric ones are possible or could be enforced via constraints. 
Importantly, achieving the low Poisson's ratio by optimizing the eigenstructure, rather than $\nu$ directly, validates the effectiveness of targeting fundamental deformation modes.

Complementing the previous example, the third test piece (\autoref{fig:array_modes}c) targeted the inverse behavior: rigidity to bulk strain and compliance to pure and simple shears (bimodal). 
The resulting linked-octagonal unit cell is reminiscent of known extremal structures explored to maximize isotropic Poisson's ratio \cite{Sigmund1994-qf}.
Experimental testing revealed stiffnesses consistent with square symmetry ($E_1/E_2 \approx 1.01$) and high Poisson's ratios ($\nu_{12} \approx 0.56$, $\nu_{21} \approx 0.52$). 
This contrasts sharply with the result for the complementary unimodal target (\autoref{fig:array_modes}b), further illustrating the optimizer's flexibility. 
The high Poisson's ratio achieved is consistent with the design goal of maximizing the bulk-to-shear moduli ratio ($K/G$), yielding values increased from the base material's $(\nu_{base} = 0.4)$.

These examples (\autoref{fig:array_modes}a-c) demonstrate that optimizing for specific eigenstructures allows FDE to realize fundamentally complementary mechanical responses (e.g., low $\nu$ vs. high $\nu$). 
Targeting the eigenstructure effectively yielded designs that minimized or maximized Poisson's ratios, achieving goals often pursued through direct design around scalar properties \cite{Andreassen2014-tc, Kadic2012-qy, Milton1992-wm} but via a more fundamental, direction-driven approach.

\subsection{Programmed Normal-Shear Coupling and Chirality}\label{sec:normal-shear}
The fourth example (\autoref{fig:array_modes}d) demonstrates FDE's ability to create chiral foams with programmed normal-shear coupled strains. 
The design target was unimodal compliance under simultaneous, similarly-signed vertical and shear strains. 
This directional coupling naturally leads to a chiral topology resembling a four-bar parallelogram linkage used in other metamaterials~\cite{Lipton2018-um}. 
Experimental testing yielded $E_1/E2 \approx 1.76$, $\nu_{12} \approx 0.30$, $\nu_{21} \approx 0.14$ (\autoref{tab:meas_sim_data}). 
Crucially, due to the handed nature of the design, non-zero normal-shear coupling terms emerge: $\eta_{212} \approx 0.72$, reflecting the design target for vertical-shear coupling, while the low value of $\eta_{112} \approx 0.03$ indicates minimal shear coupling when strained horizontally, consistent with the design's aim to resist deformations other than the primary vertical-shear mode.

\begin{figure}[htbp]
    \centering
    \includegraphics[width=\linewidth]{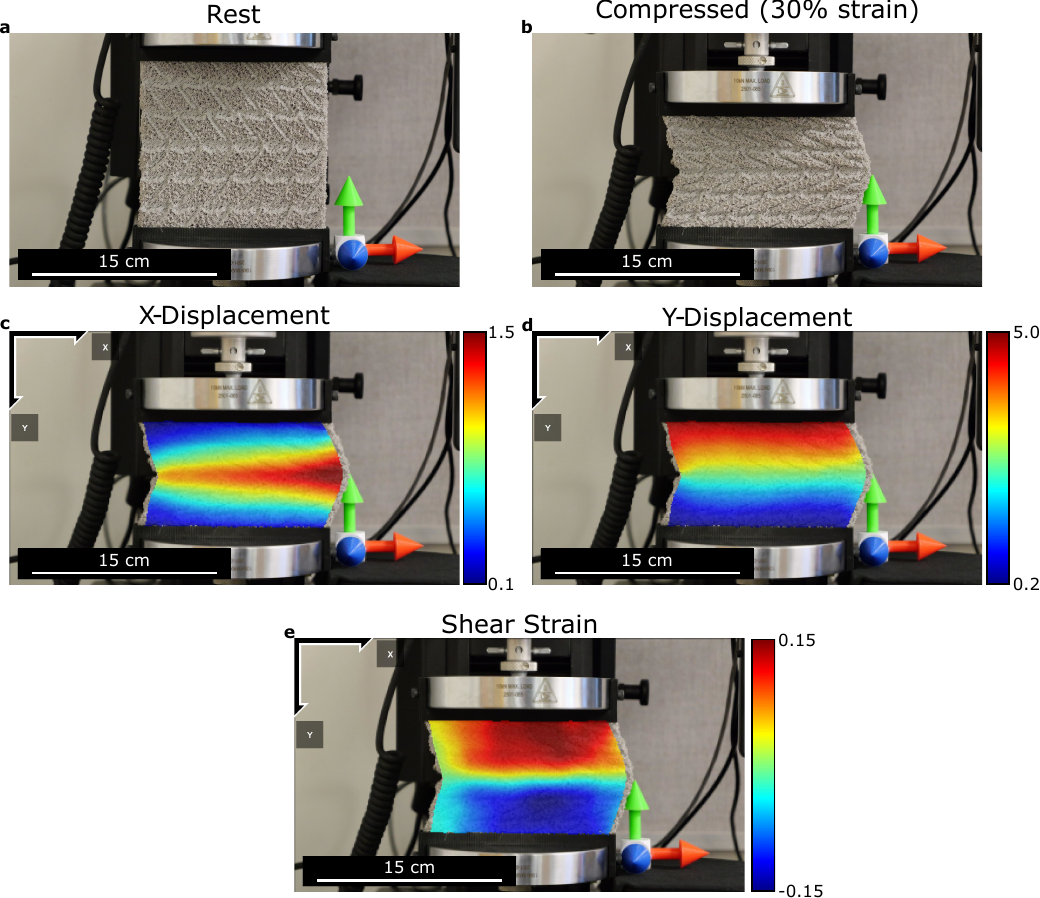}
    \caption{DIC analysis of a compression test of the handed shearing material (\autoref{fig:array_modes}c). The foam is mirrored about the horizontal axis to prevent shear stresses at the test boundaries. a-b) The material at rest and 30\% compressive strain (45 mm of compression). c) X-Displacement of the foam showing the center of the material displacing to the right. The banded nature indicates how the material shifts uniformly on top and bottom. d) Y-Displacement of the foam showing uniform vertical compression. e) The tensorial Eulerian-Almansi shear strain ($\varepsilon_{xy}=\gamma_{xy}/2$) showing symmetric shear strain mirrored about the middle. 
    }
    \label{fig:DIC}
\end{figure}

The targeted vertical-shear coupling is clearly visualized via digital image correlation (DIC) during compression testing, shown in \autoref{fig:DIC}. 
The foam, compressed vertically ($x_2$), exhibits a clear shear deformation coupled with the compression, visible in the x- and y-displacement fields (\autoref{fig:DIC}c-d) and the Euler-Almansi shear strain field (\autoref{fig:DIC}e). 
Note, the tested sample was mirrored horizontally to mitigate boundary effects, thus the shear strains seen in \autoref{fig:DIC}e have opposite signs from top to bottom. 
The measured relatively low Poisson's ratio ($\nu_{21}\approx0.14$) confirms that the observed horizontal displacement is dominated by the programmed normal-shear coupling, not the Poisson effect.

To our knowledge, this represents the first demonstration of programmed chirality---arising directly from encoding handed topologies via FDE---within monolithic foam materials built from a single continuous filament. 
Distinct from periodic lattice-based chiral metamaterials~\cite{Alderson2010-nc} or foams exhibiting merely statistical chirality, achieving these significant normal-shear coupling terms ($\eta_{212}\approx0.72$,~\autoref{tab:meas_sim_data}) purely through patterned mesostructure highlights a unique capability of FDE. 
This ability to program chirality expands the functional possibilities for entangled materials, moving beyond previously realized isotropic or simple anisotropic responses and suggesting potential uses in transducers or systems requiring engineered extension under shear/torsional load or vice versa.

Collectively, these fabricated examples validate that FDE, by manipulating local entanglement based on targeted eigenstructures, allows the creation of monolithic foams exhibiting a range of tailored mechanical behaviors---including anisotropy and programmed chirality---that expand upon the capabilities of conventional entangled materials.

\subsection{Comparison with Simulation}\label{sec:compare_sim}
Comparing the measured initial effective properties with the fabrication-calibrated simulations (\autoref{tab:meas_sim_data}) reveals general agreement, validating the FDE approach's ability to achieve targeted mechanical behaviors. 
When accounting for fabrication minimum feature size, the simulations accurately capture the primary stiffness responses (Young's moduli $E_1,~E_2$) and crucially predict the presence or absence of significant normal-shear coupling ($\eta_{212}$) consistent with the chiral or symmetric design intent.
Accounting for the fabrication minimum feature size required calibrating simulation parameters; additional details on the calibration methodology are given in \autoref{method:calibration} and in the SI section S3. 

Quantitative agreement is weaker for other secondary properties, notably the Poisson's ratios ($\nu_{12},~\nu_{21}$) in certain cases (e.g., the Unimode Bulk design).
Discrepancies also appear in lower normal-shear coupling terms.
For instance, the simulated $\eta_{112}$ (0.14) differs notably from the measured value (0.03), potentially influence by implementing the horizontal mirroring applied during fabrication to manage test boundary conditions.
Furthermore, small non-zero shear coupling terms (e.g., $\eta_{112}\approx0.01-0.04$) were measured for the nominally symmetric designs where simulations predict zero. 
We hypothesize these minor terms arise from toolpath orientation effects relative to the testing axes, which are perhaps incompletely canceled by the layer-by-layer stacking strategy. 
Overall, remaining discrepancies likely reflect the combined influence of inherent VTP variability, simplifications in the linear elastic material model, and geometry inaccuracies tied to the post-calibration step.

Therefore, this assessment suggests that when interpreting the simulated property space in \autoref{fig:mat_props}, the results for Young's moduli ($E_1,~E_2$) and major normal-shear coupling effects (e.g., significant $\eta_{212}$) provide a good indication of achievable quantitative trends via FDE.
Conversely, predictions involving Poisson's ratios ($\nu_{ij})$ and minor coupling terms should be viewed more qualitatively, reflecting the greater simulation-experimental discrepancies observed for these properties.

\subsection{Expanded Material Property Space}\label{sec:expanded_mat_props}
\begin{figure}[htbp]
    \centering
    \includegraphics[width=\linewidth]{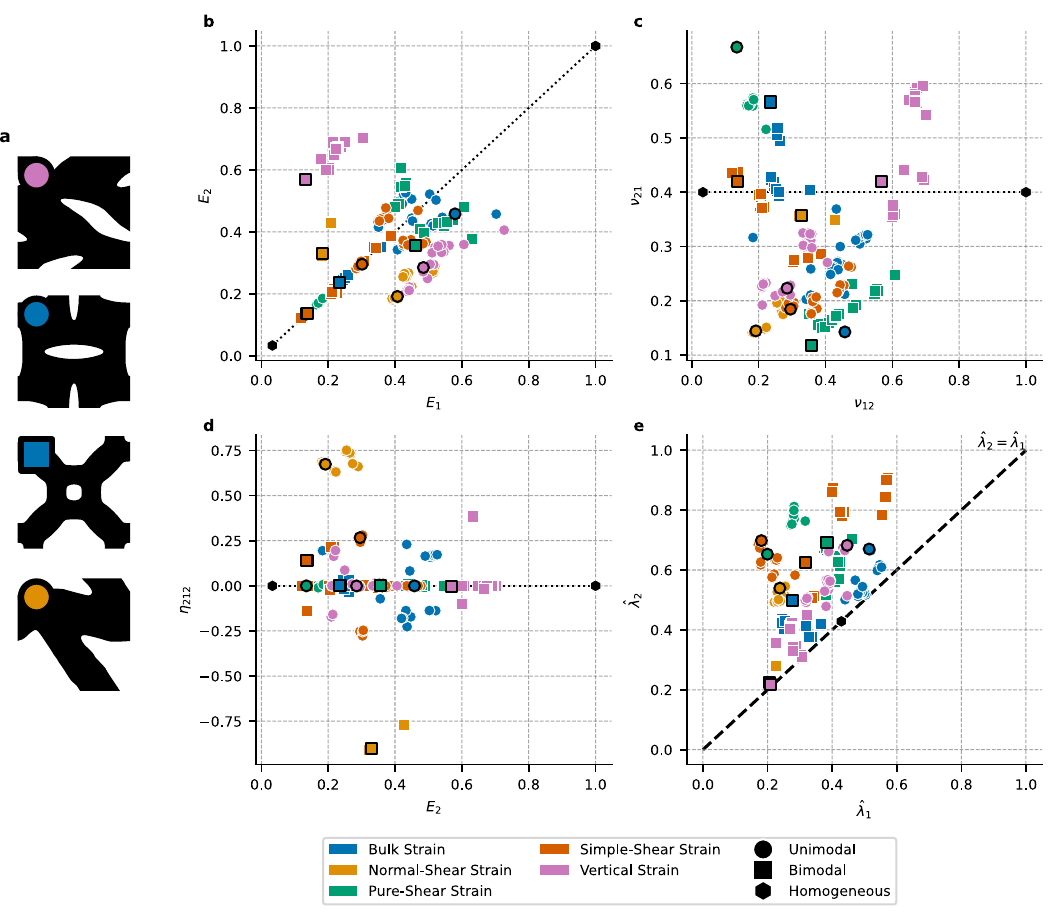}
    \caption{Simulated material parameter mappings from topology optimization runs (n=247), highlighting the expanded material property space achievable via FDE compared to homogeneous foams. Hexagonal black points indicate homogeneous foam samples, connected by a line representing linear interpolation. a) Marker legend identifying the four topology-calibrated designs from Fig. 2 denoted with a black marker border on the plots. The remaining topologies serve as illustrative examples of the other design families. b) Comparison of Young's modulus in principal directions ($E_2$ vs $E_1$), demonstrating one form of anisotropy. c) The vertical stiffness and resulting Poisson's ratio ($\nu_{21}$ vs. $E_2$), showing the expanded range of values compared to the approximated homogeneous baseline (constant $\nu=0.4$). d) The vertical stiffness and resulting normal-shear coupling term ($\eta_{212}$ vs. $E_2$), illustrating the creation of materials with positive and negative normal-shear coupling. e) Normalized eigenvalues of the elasticity matrix $C$, providing a coordinate-independent representation of the materials' principal stiffness magnitudes and overall anisotropy.}
    \label{fig:mat_props}
\end{figure}

\autoref{fig:mat_props} presents the expanded material property space predicted by our calibrated simulations, which incorporate corrections for fabrication resolution limits as detailed in the Methods (\autoref{method:calibration}) and Supplemental Information (section S3).
A key insight is that the FDE-patterned foams occupy regions unattainable by this naive blending, highlighting the expanded design freedom.
The results (n=247 simulations) are summarized in \autoref{fig:mat_props}, plotting various effective property combinations. 
Each symbol type corresponds to a prescribed mode (circle is unimode, square is bimode), while colors indicate the targeted strain direction for compliance or rigidity. 
The four fabricated designs (\autoref{fig:array_modes}) and additional designs are marked with black borders (\autoref{fig:mat_props}a).
The black hexagons representing the homogeneous sparse and dense foams, and the dashed line connecting them represents a simple linear interpolation (mixing rule). 
A limitation of analyzing traditional material properties in plots \autoref{fig:mat_props}b-d is their implicit dependence on a coordinate system, potentially obfuscating material similarities. 
We address this in \autoref{fig:mat_props}e by analyzing normalized eigenvalues---a coordinate- and stiffness-invariant metric---to investigate those relationships.

\autoref{fig:mat_props}b compares principal moduli $E_1$ and $E_2$ to illustrate directional stiffness with points along the diagonal ($E_1=E_2$) representing isotropic/square-symmetric responses. 
While many FDE designs cluster near this diagonal, several families exhibit distinct anisotropy consistent with design targets. 
For instance, vertically compliant unimodes (pink circles) lie below the diagonal ($E_1>E_2$), while their complementary bimodes (pink squares) lie above it ($E_1<E_2$). 
Normal-shear designs (yellow markers) also show this reflection across the diagonal based on their mode. 
Furthermore, several design families (e.g., pure-shear, simple-shear, bulk strain) contain solutions both on or near the diagonal ($E_1\approx E_2$) and away from it ($E_1 \ne E_2$), corresponding to the optimizer finding either symmetric or asymmetric design rotated by 90\degree. 
The existence of both symmetric and asymmetric outcomes for the same functional target demonstrates that FDE can access diverse topologies without being restricted to a single symmetry type. 
\autoref{fig:mat_props}b thus indicates that despite the constraints of FDE (e.g., moderate stiffness range), the optimization framework can target distinct anisotropy, yielding predicted directional stiffness different from homogeneous base materials, while also revealing potential symmetries in the optimized topologies.

\autoref{fig:mat_props}c plots Poisson’s ratio $\nu_{21}$ against vertical modulus $E_2$. 
The FDE designs show a wide distribution, expanding beyond the base material value ($\nu = 0.4$). 
This demonstrates that FDE optimization not only broadens the range of available Poisson's ratios but also allows achieving these varied ratios across a considerable range of vertical stiffness ($E_2$ spanning approximately 0.15 to 0.70). 
For instance, the families targeting pure-shear strain (green markers) appear particularly effective at reaching the highest and lowest Poisson's ratios (i.e., high and low $K/G$ respectively) in this coordinate system, while the bulk strain family (blue markers), despite also influencing K/G, does not reach the same extremes. 
This difference can likely be attributed to the implicit encoding of symmetry in the formulation of $T^*$ where $V_{pure-shear}$ encodes square symmetry while $V_{bulk}$ encodes the more restrictive isotropic symmetry. 
It follows that the choice of target eigenstructure, particularly the coordinate system and symmetry defined by $V$, directly influences the achievable range of properties like Poisson's ratio within the FDE framework.

\autoref{fig:mat_props}d plots the normal-shear coupling coefficient $\eta_{212}$ against $E_2$, quantifying chirality. 
As intended, designs explicitly targeting normal-shear coupling (yellow markers) exhibit distinct positive (unimode) or negative (bimode) $\eta_{212}$ values, clearly separated from zero and generally larger in magnitude than the incidental coupling seen in other families that do not target the coupling.
This apparent coupling can arise from two distinct mechanisms: either the optimizer finds a truly topologically chiral solution lacking mirror symmetry, or a mirror-symmetric structure is measured in a coordinate system rotated away from its principal axes of symmetry. 
Both mechanisms are observed here. 
For example, the simple-shear targets (orange markers) produced both topologically chiral solutions (e.g., \autoref{fig:mat_props}a unimode example) and mirror-symmetric structures (e.g., the rotated ``inverted honeycomb’’ topology in \autoref{fig:mat_props}a bimode example) that exhibit coupling in this coordinate frame. 
Similarly, the unimode bulk strain family (blue circles), aiming to minimize $K/G$ (equivalent to minimizing Poisson's ratio), often converged to topologically chiral solutions, consistent with established links between chirality and auxetic behavior~\cite{Alderson2010-nc}. 
Overall, the ability to achieve both targeted positive and negative coupling, alongside the emergence of incidental chirality (whether topological or coordinate-system-based), confirms FDE allows tuning of the handed response and provides multiple mechanisms for generating structures exhibiting normal-shear coupling.

\autoref{fig:mat_props}e provides a scale- and coordinate-invariant perspective using normalized eigenvalues, illuminating the fundamental range of intrinsic material behaviors that can be accessed by FDE.
If \(\lambda_1 \le \lambda_2 \le \lambda_3\) are the eigenvalues of a design’s elasticity matrix \(C\), we project the materials behavior onto the pair \(\bigl(\hat{\lambda}_1, \hat{\lambda}_2\bigr)\), where \(\hat{\lambda}_i = \lambda_i/\lambda_3 \quad (i=1,2,3)
\).
In this space, both homogeneous base materials collapse to a single point (0.43, 0.43), as they share the same Poisson's ratio. 
The FDE designs, however, scatter across a wide area, visually demonstrating the expansion of the intrinsic property space enabled by patterned entanglement.

This plot reveals underlying similarities obscured in other views; for instance, families spanning wide property ranges in other plots (e.g., unimode bulk-strain, blue circles) cluster more tightly, suggesting similar intrinsic stiffness ratios of $C$ are achieved and that the spread in other plots is due to differences in volume fraction or eigenvector misalignment between $V$ and $T^*$. 
Furthermore, overlaps between families targeting related deformations suggest topological similarity; for example, the simple- and pure-shear unimode designs (orange and green circles) cluster here, reflecting their linked mechanics and similar topologies (differing mainly by a 45\degree rotation, see~\autoref{fig:mat_props}a). 
The tightness of clustering also varies (e.g., tight normal-shear clusters vs. broader bulk-strain clusters), hinting at the relative ease or difficulty of achieving certain targets and the potential for multiple distinct and dominating topological solutions. 
Thus, the substantial spread of FDE results compared to the single homogeneous point, along with the varied clustering and overlaps, underscores FDE's ability to reshape stiffness distributions and access diverse intrinsic material behaviors, offering multiple design pathways to achieve specific mechanical objectives, even with moderate material contrast.

Overall, the simulation results (\autoref{fig:mat_props}) indicate that FDE-based foams occupy a far richer design space than uniform blends. 
The ability to tune anisotropy, Poisson's ratio, and particularly normal-shear coupling underscores that, when properly controlled, entangled materials can have emergent properties not seen previously.
While the $\sim30{:}1$ stiffness contrast via VTP precludes finding truly extremal material properties at the bounds of elastic theory, the use of TO targeting specific eigenstructures (inspired by extremal materials theory) is designed to find optimal material arrangements.
These optimized structures, achieving diverse and complex behaviors (\autoref{fig:mat_props}), serve as computational probes exploring the performance limits imposed by the finite material contrast.

\section{Conclusion}
This work introduces and validates Fabrication-Directed Entanglement (FDE), a methodology integrating topology optimization (TO) with viscous thread printing (VTP), to unlock spatially programmed mechanical properties in monolithic foam materials.
By computationally guiding VTP deposition to create quasi-two-phase (dense/sparse) regions, we successfully implemented and experimentally validated an FDE workflow. 
Fabricated prototypes realized targeted anisotropic responses, tuned Poisson's ratios, and, critically, demonstrated chiral foam structures exhibiting significant programmed normal-shear coupling.
Simulation results further illustrated that FDE broadens the accessible property space for VTP-based materials compared to uniform structures.
By targeting specific eigenstructures inspired by extremal materials theory, the optimization framework finds tailored mesostructures that serve to explore the boundaries of this achievable space, making effective use of the VTP system's $\sim30{:}1$ contrast.

FDE provides a new capability for designing and fabricating hierarchical foam mechanical metamaterials from a single homogeneous filament. 
The key significance lies in demonstrating that this approach can unlock complex mechanical behaviors previously inaccessible in such monolithic foam structures, notably the patterned chirality and resultant normal-shear coupling. 
Achieving these handed responses purely through patterned mesostructure showcases the power of combining computational design with controlled VTP deposition.
This integrated method provides a route to functional materials leveraging both designed mechanics and the inherent advantages of a continuous, monolithic build process, such as global network connectivity.

While FDE enables complex programmed mechanics in monolithic foams, key trade-offs arise from the underlying VTP process. 
The achievable stiffness contrast ($\sim30{:}1$), while sufficient to demonstrate behaviors like chirality, restricts the property range compared to high-contrast TO methods and limits access to theoretically extremal materials. 
Similarly, although the continuous filament path ensures inherent network connectivity---a potential advantage over discrete fabrication approaches---the process resolution ($\sim1.2~\text{mm}$ coil diameter) fundamentally limits fine feature realization and poses challenges for miniaturization. 
Additionally, the current reliance on post-fabrication calibration (\autoref{method:calibration}, section S3), necessitated by resolution-limited feature fidelity (i.e., effective feature thickening) at the demonstrated scale, currently limits the ability to reliably predict the quantitative properties of new FDE designs entirely a priori, necessitating further validation or potentially iterative calibration.
Finally, stability considerations, highlighted by the early buckling ($\sim5\%$ strain) observed in the tested high-aspect ratio geometries, suggest that practical applications in unconfined compression may require operation at low strains or designs optimized for greater stability under load.

Looking ahead, future efforts should focus on both enhancing the current FDE methodology and expanding its scope. 
Immediate priorities include improving fabrication fidelity and developing more predictive simulation models that better capture VTP fabrication limitations a priori, thereby reducing reliance on post-hoc calibration and enhancing design accuracy. 
Alongside these refinements, exploring methods to increase the achievable stiffness contrast could push material properties closer to theoretical limits. 
Furthermore, a crucial next step involves extending the FDE framework to three dimensions to unlock its full potential for complex component design. 
Beyond refining the core method, the demonstrated control over Cauchy elasticity opens avenues for incorporating dynamics (e.g., targeting elastic bandgaps) or extending the chiral work to incorporate Cosserat elasticity. 
Finally, fully characterizing these hierarchical materials necessitates investigating potential entanglement-related properties, such as damping and toughness, in conjunction with the programmed elastic response explored here.

\section{Methods}\label{sec:methods}
\subsection{Materials \& Fabrication}
\label{method:mat_and_fab}
We fabricated the samples using NinjaTek NinjaFlex\textregistered{} thermoplastic polyurethane (TPU) (density 1.19 g/cc, tensile modulus 12 MPa, 85 Shore A) as the base material stored in and printed from a filament dryer. 
The extrusion temperature was 240\degree C, bed temperature 70\degree C, on a Prusa Mk4 fused filament fabrication (FFF) printer with a 0.4 mm nozzle.
The samples were printed with the plane of the pattern being the \(XY\)-plane of the printer; however, with the toolpath traversing 45\degree to the sample principal axes.
We used the Gaussian process regression model from \cite{Emery2024-bb} to select two \vhstar{} pairs for low-density (LD, \((0.4, 14.81)\)) and high-density (HD, \((0.15, 6.93)\)) regions with line spacing \((dL)\) 1.2 mm and layer height \((dz)\) 1.27 mm.

Two types of samples were fabricated: dense and sparse homogeneous cubes ($\sim40~\text{mm}$ side length) using the specified LD/HD parameters, and $6\times 6$ metamaterial unit cell arrays yielding samples size $\sim150\times150\times28~\text{mm}^3$.
The cubic samples took approximately 15 hours to print and the metamaterial samples took approximately 40 hours to print.
Preparing the metamaterial designs involved projecting the density fields from IH into binary layouts using a filter and threshold approach with TO projection parameters $\beta=64$ (projection strength) and $\chi=0.5$ (dilation factor) based on previous work~\cite{Wang2011-gy, Andreassen2014-tc}.
These resulting binary bitmaps were then converted into CAD files for slicing and fabrication.
Both types of samples were then processed with a custom slicer to generate the VTP specific G-Code~\cite{Nakura2024-yp}.

\subsection{Modulus Measurement}
Both homogeneous and metamaterial samples were tested under uniaxial compression using an Instron 6844 universal testing machine, capturing force-displacement data and simultaneous time-lapse images (1 frame/sec) for subsequent DIC analysis (see~\autoref{method:DIC}).
Effective stress and engineering strain were calculated using initial sample dimensions. 
Homogeneous cubic samples exhibited linear stress-strain behavior up to 10\% strain and were tested accordingly to determine their Young's modulus ($E_{HD}\approx3.4~\text{MPa},~E_{LD}\approx0.11~\text{MPa}$) via linear fits over this range. 
Metamaterial samples were also tested up to 10\% strain, but buckling became apparent above approximately 5\% strain.
To avoid confounding effects from this instability, the effective properties for metamaterials (summarized in~\autoref{tab:meas_sim_data}) were extracted solely from the initial 0-2.5\% strain region.
Stress-strain curves confirm this region remained appropriately linear for material property estimation.

\subsection{Digital Image Correlation}
\label{method:DIC}
Digital Image Correlation (DIC) was used to measure displacement and strain fields during compression testing using the freely available MATLAB package Ncorr \cite{Blaber2015-gt}. 
For the homogeneous samples the effective Poisson's ratios were measured as $\nu_{LD}\approx0.45$, $\nu_{HD}\approx0.35$ in the same 0-10\% strain range as the modulus.
The metamaterial properties were measured in the 0-2.5\% strain range, again due to buckling observed at higher strains, and are summarized in~\autoref{tab:meas_sim_data}. 
Implementation details given in section S1 of Supplemental Information.

\subsection{Computational Methods}
Finite element simulations using periodic boundary conditions were implemented in FEniCS~\cite{Alnaes2015-vr} on a 2D cross-diagonal triangular mesh (unit cell domain side length = 1). 
Periodic homogenization was applied to extract the effective elasticity matrix $C$~\cite{Andreassen2014-en, Sigmund1994-qf}. 
Optimization was performed using NLopt~\cite{JohnsonUnknown-pt} using the Method of Moving Asymptotes (MMA)~\cite{Svanberg1987-dd}.


\subsection{Calibration of Simulation Parameters}\label{method:calibration}
Initial comparisons between simulated and experimentally measured effective properties revealed discrepancies. 
These were primarily attributed to the resolution limits inherent in the VTP process, which effectively thickens fine features in the fabricated samples compared to the idealized topology optimization outputs. 
To improve the agreement and ensure simulations accurately reflected the mechanical behavior of the fabricated structures, we performed an experimental calibration.

This calibration involved optimizing key simulation parameters to minimize the difference between simulated and measured mechanical properties (specifically $E_1$, $E_2$, $\nu_{12}$, and $\nu_{21}$). 
Through this process, the topology optimization projection dilation factor ($\chi$) was identified as the dominant parameter influencing the agreement, as it directly compensates for the fabrication-induced feature thickening. 
All simulated properties reported subsequently in this work (\autoref{tab:meas_sim_data}, \autoref{fig:mat_props}) utilize the calibrated dilation factor of $\chi=0.15$ (compared to the initial value of $\chi=0.5$).

Further details on the calibration methodology, including the objective function and the complete set of optimized parameters, along with a comprehensive comparison of simulated properties before and after calibration, can be found in the Supplemental Information section S3.

\backmatter

\bmhead{Acknowledgments}
\textbf{Funding}:This work was funded by the National Science Foundation grant number IIS-2212049, a gift from Ford Motor Company, a grant from The Murdock Charitable Trust.
\textbf{Competing Interests:} The authors have a patent application for path planning methods used (Patent Application 63/527,296 filed 7/17/2023).
\textbf{Author Contributions:} D.R. developed the software,performed the mechanical testing, and did the data analysis. J.I.L. and D.R. conceived of the work and wrote the paper. J.I.L provided funding. 
\textbf{Data and materials availability:} All data and source code are publicly available at https://github.com/TransformativeRoboticsLab/FDE-Metamaterials.
The data is also accessible at https://bit.ly/TO-VTP.




\noindent
\bibliography{paperpile}

\end{document}


\maketitle

\section{Digital Image Correlation (DIC)}\label{sec:DIC}
Digital Image Correlation (DIC) was used to measure displacement and strain fields during compression testing using the freely available MATLAB package Ncorr \cite{Blaber2015-gt}.
Images were taken approximately every one second over a compression range of 0-10\%.
The homogeneous cube samples were characterized in the full 0-10\% linear strain range, similar to modulus measurements.
The metamaterial samples were characterized in the linear 0-2.5\% strain range due to buckling observed at higher strains.

\begin{table}[htbp]
  \centering
  \caption{DIC Parameters}
  \label{tab:dic_params} 
  \begin{tabular}{lc}
    \toprule
    Parameter                 & Value    \\
    \midrule 
    Radius                    & 20       \\
    Strain Radius             & 5        \\
    Subset Spacing            & 4        \\
    Norm Cutoff               & 1e-6     \\
    Iteration Cutoff          & 100      \\
    DIC Subset Truncation     & Disabled \\
    Strain Subset Truncation  & Disabled \\
    \bottomrule
  \end{tabular}
\end{table}

\begin{figure}[htbp]
    \centering
    \includegraphics[width=0.9\linewidth]{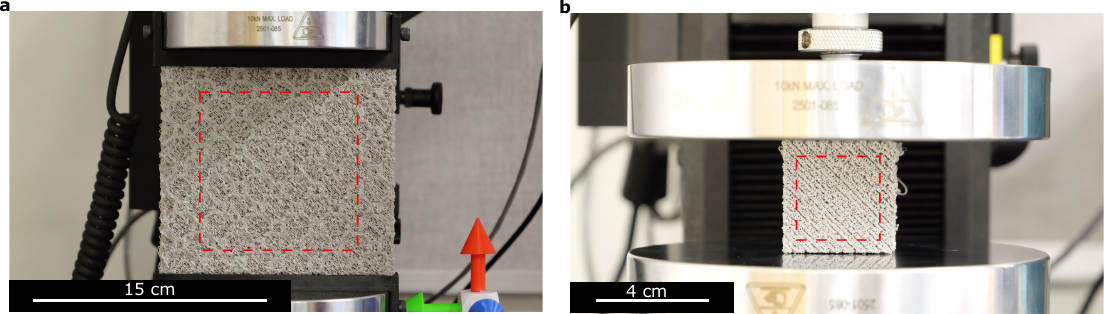}
    \caption{The approximate ROI placement used for the metamaterial (a) and homogeneous (b) samples.}
    \label{fig:DIC_ROI}
\end{figure}

A region of interest (ROI) was selected covering the inner portion of the sample face (e.g., central 4×4 unit cell region for the 6×6 metamaterial arrays; similar ratio for homogeneous cubes) to minimize boundary effects (see~\autoref{fig:DIC_ROI}).
For DIC analysis, the coordinate system was defined with the positive $y$-direction aligned with compressive loading and positive $x$-direction toward the right (see main text Figure 3c-d).
To determine directional material properties such as $E_1$ vs. $E_2$ the material axes ($x_1$,$x_2$) were mapped based on sample orientation.

First, the full discretized Lagrangian strain fields $\varepsilon_x(x_i,y_i,t)$, $\varepsilon_y(x_i,y_i,t)$, and $\varepsilon_{xy}(x_i,y_i,t)$ were determined for the ROI using the parameters in~\autoref{tab:dic_params}. 
At each time step $t$, homogenized strain values were calculated by averaging the respective strain fields over the Region of Interest (ROI) for all $N$ points in the ROI: 
\[
\begin{aligned}
\bar{\varepsilon}_x(t) &= \frac{1}{N}\sum_{i=1}^N \varepsilon_x(x_i,y_i,t) \\
\bar{\varepsilon}_y(t) &= \frac{1}{N}\sum_{i=1}^N \varepsilon_y(x_i,y_i,t) \\ \bar{\varepsilon}_{xy}(t) &= \frac{1}{N}\sum_{i=1}^N \varepsilon_{xy}(x_i,y_i,t).
\end{aligned}
\]

\begin{figure}
    \centering
    \includegraphics[width=\linewidth]{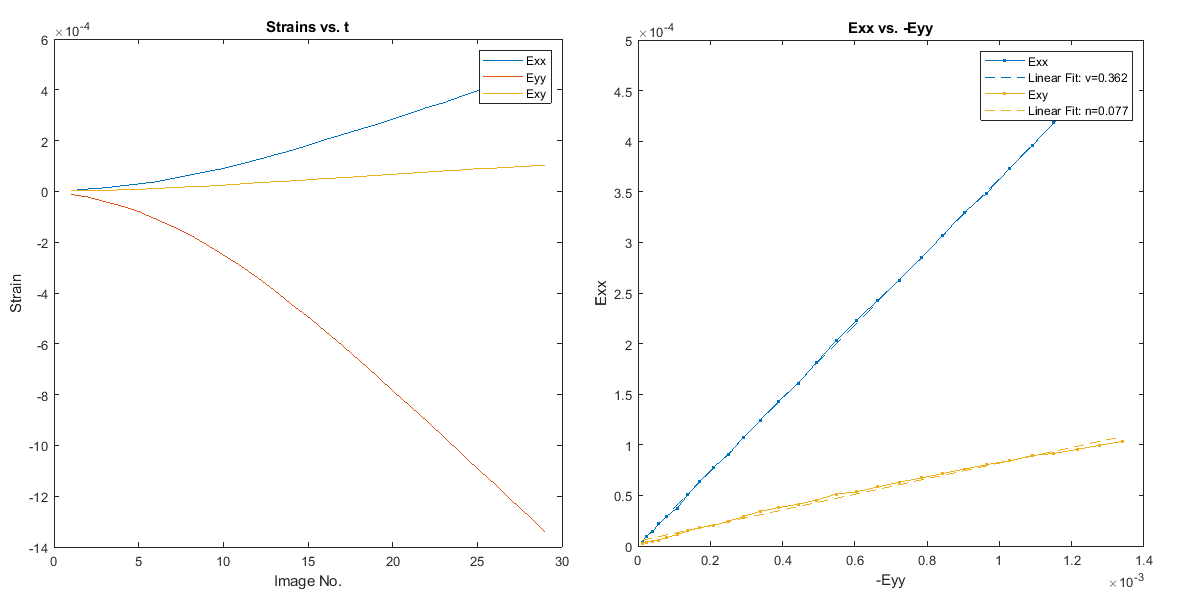}
    \caption{Example for curve fitting of time-dependent strain values.}
    \label{fig:strain_plot}
\end{figure}

The effective Poisson's ratio $\bar{\nu}_{yx}$ was determined by performing a linear fit on the ROI-averaged strain data, plotting $\bar{\varepsilon}_x(t)$ against $\bar{\varepsilon}_y(t)$ for the range where $\bar{\varepsilon}_y(t)$ is between 0\% and 2.5\%. The effective Poisson's ratio is the negative slope of this fit: $\bar{\nu}_{yx} = -d\bar{\varepsilon}_x / d\bar{\varepsilon}_y$.
An example for the dense homogeneous samples is shown in~\autoref{fig:strain_plot}.

The normal-shear coupling coefficient $\bar{\eta}_{yxy}$ was calculated similarly for non-chiral samples. 
A linear fit was performed on the $\bar{\varepsilon}_{xy}(t)$ vs. $\bar{\varepsilon}_y(t)$ data over the same 0-2.5\% strain range for $\bar{\varepsilon}_y(t)$. 
The slope gives the coefficient: $\bar{\eta}_{yxy} = d\bar{\varepsilon}_{xy} / d\bar{\varepsilon}_y$. 
Non-zero values occurred for non-chiral samples (Figure 2); however, this was due to the 45\degree print angle of the samples rather than any significant shear effects.
Note this definition gives half of the engineering strain $d\bar\gamma_{xy}/d\bar{\varepsilon}_y$ definition. 
See~\autoref{fig:strain_plot} for illustration of linear fitting.

For the mirrored chiral sample (main text Figure 2d), a modified approach was needed because the mirroring of the array results in opposite signed shear strains, thus meaning the na\"ively averaged shear strain in the ROI would zero out.
While the ROI is approximately centered on the array, the ROI midline may not exactly align with the sample midline.
Thus, we located the sample midline in ROI coordinates $y_{mid}$ by averaging the rows of shear strain and finding the index of the minimum value closest to the center of the ROI: $\min |\text{mean}(\varepsilon_{xy}(x,y_{row},t))|$. 
Then, to account for the mirror symmetry, a modified shear strain field $\varepsilon_{xy,mod}$ was created by multiplying the shear strain field values below the midline by -1: 
\[
\varepsilon_{xy,mod}(x,y,t) =
\begin{cases}
  \varepsilon_{xy}(x,y,t) & \text{if } y \ge y_{mid} \\
  -\varepsilon_{xy}(x,y,t) & \text{if } y < y_{mid}.
\end{cases}
\]
 The ROI-averaged modified shear strain, $\bar{\varepsilon}_{xy,mod}(t)$, was calculated by averaging the $\varepsilon_{xy,mod}(x,y,t)$ field over the ROI. 
 Finally, a linear fit was performed on the $\bar{\varepsilon}_{xy,mod}(t)$ vs. $\bar{\varepsilon}_y(t)$ data (over the 0-2.5\% strain range for $\bar{\varepsilon}_y(t)$). 
 The slope of this fit yielded the normal-shear coupling coefficient $\bar{\eta}_{yxy}$ for this sample.

\section{FDE Inverse Homogenization}
Our IH approach employs standard density-based TO informed by extremal material theory. 
The target eigenstructure $(V,\Lambda)$ is user-defined: $V$ provides the orthonormal basis for the desired principal directions (\autoref{tab:V_list}), and $\Lambda=\text{diag}([a, 1, 1])$ sets the relative eigenvalue ratio with $a \ll 1$. 

\begin{table}[htbp]
    \centering
    \caption{The list of orthonormal $V$ matrices used in this work to construct the optimization target $T^*=V \Lambda V^T$.}
    \label{tab:V_list}
    \begin{tabular}{lccccc}
        \toprule
        Lead Dir. & Bulk & Normal-Shear & Pure-Shear & Simple-Shear & Vertical \\
        \midrule
        $V$ &
        $\begin{pmatrix} 
            \isqrt{1}{2} & \isqrt{1}{2} & 0 \\ 
            \isqrt{1}{2} & \isqrt{-1}{2} & 0 \\ 
            0 & 0 & 1 
        \end{pmatrix}$ &
        $\begin{pmatrix} 
            0 & 0 & 1 \\ 
            \isqrt{1}{2} & \isqrt{-1}{2} & 0 \\ 
            \isqrt{1}{2} & \isqrt{1}{2} & 0 
        \end{pmatrix}$ &
        $\begin{pmatrix} 
            \isqrt{1}{2} & \isqrt{1}{2} & 0 \\ 
            \isqrt{-1}{2} & \isqrt{1}{2} & 0 \\ 
            0 & 0 & 1 
        \end{pmatrix}$ &
        $\begin{pmatrix} 
            0 & 0 & 1 \\ 
            0 & 1 & 0 \\ 
            1 & 0 & 0 
        \end{pmatrix}$ &
        $\begin{pmatrix} 
            0 & 1 & 0 \\ 
            1 & 0 & 0 \\ 
            0 & 0 & 1 
        \end{pmatrix}$ \\
        \bottomrule
    \end{tabular}
\end{table}

Based on the measured stiffness contrast between dense and sparse homogeneous VTP foams, the corresponding normalized stiffness values used for material interpolation within TO were $E_{max}=E(\rho=1)=1=E_{HD}/E_{HD}$ and $E_{min}=E(\rho=0)=1/30\approx E_{LD}/E_{HD}$. 
We used the minimum stiffness and an approximate lower bound to set the target eigenvalue ratio $a = 1/30$. 
An effective Poisson's ratio $\nu=0.4$ was assumed for both phases, representing an intermediate value between the measured $\nu_{LD}\approx0.45$ and $\nu_{HD}\approx0.35$. 
While standard TO constraints (e.g., volume fraction) can be included ($g_n(\rho)$), they were found unnecessary to achieve distinct, well-connected designs with this FDE framework.


\section{Post-Fabrication Calibration}\label{sec:calibration}
This section provides detailed information regarding the post-fabrication calibration process outlined in the main text (section 6.5). Comparisons between initial finite element simulations and experimental mechanical tests revealed discrepancies in the predicted effective properties of the fabricated metamaterials. As discussed in the main text, these differences are primarily attributed to the resolution limitations inherent in the VTP process, which results in the effective thickening of the thinnest features in the physical samples compared to the idealized designs generated by TO.

To reconcile these results and improve the predictive capability of the simulations, an experimental calibration was performed. 
This involved a parametric optimization against key factors influencing the simulation output: the TO base material properties ($E_{max}$, $E_{min}$, $\nu$) and TO projection parameters ($\beta$, $\chi$). 
The optimization was set up to minimize a weighted sum of the squared normalized errors between simulated and measured homogenized properties. 
The properties included in the calibration were $E_1$, $E_2$, $\nu_{12}$, and $\nu_{21}$. 
Weights were applied to the squared normalized error of each property to balance their influence in the objective function: a weight of 1 was used for the errors in the Young's moduli ($E_1$, $E_2$), and a weight of 0.1 was used for the errors in the Poisson's ratios ($\nu_{12}$, $\nu_{21}$). 
The objective function is defined as:
$$\min_{P \in \{E_{max}, E_{min}, \nu, \beta, \chi\}} \sum_{\text{designs}} \sum_{p \in \{E_1, E_2, \nu_{12}, \nu_{21}\}} w_p \left(\frac{p_{\text{sim}}(P) - p_{\text{measured}}}{p_{\text{measured}}}\right)^2$$
where $w_p = 1$ for $p \in \{E_1, E_2\}$ and $w_p = 0.1$ for $p \in \{\nu_{12}, \nu_{21}\}$.

Normal-shear coupling is a material property characteristic of chiral structures. Given that the majority of the designs included in this calibration optimization were optimized for symmetric responses (e.g., uniaxial, bulk, or pure shear compliance/rigidity, as shown in main text Figure 2a-c), their ideal, intended behavior exhibits zero normal-shear coupling due to symmetry. 
Consequently, including $\eta$ terms, which are only significant for explicitly chiral designs (main text Figure 2d), in a global calibration objective pooling data from all designs was deemed inappropriate. 
The calibration was thus focused on optimizing parameters to improve the agreement for properties ($E_1$, $E_2$, $\nu_{12}$, $\nu_{21}$) relevant to all design symmetries considered. 
Despite their exclusion from the optimization objective, reasonable agreement was still observed between the measured and calibrated simulated $\eta$ values across all designs, as shown in Table 1 of the main text.

The optimization process identified the projection dilation factor ($\chi$) as the dominant parameter significantly influencing the agreement between simulated and measured properties, with other parameters having only a marginal effect (referred to as $\eta$ in \cite{Bendsoe1988-hx, Wang2011-gy}). 
This finding aligns with the physical expectation that adjusting $\chi$ directly compensates for the fabrication-induced feature thickening by modifying the thickness of features within the simulation model. 
An error-minimizing optimal value of $\chi=0.15$ was determined, a significant adjustment from the initial value of $\chi=0.5$ used in preliminary simulations.

The effect of this calibration on the resulting topologies is illustrated in \autoref{tab:dilation}, which compares the original outputs of the TO process ($\chi=0.5$) with the post-calibrated topologies ($\chi=0.15$). 
The visual differences highlight how the adjustment of the dilation factor modifies the feature thickness in the simulation, aiming to better represent the fabricated structures.

\begin{figure}[h]
    \centering
    \includegraphics[width=\linewidth]{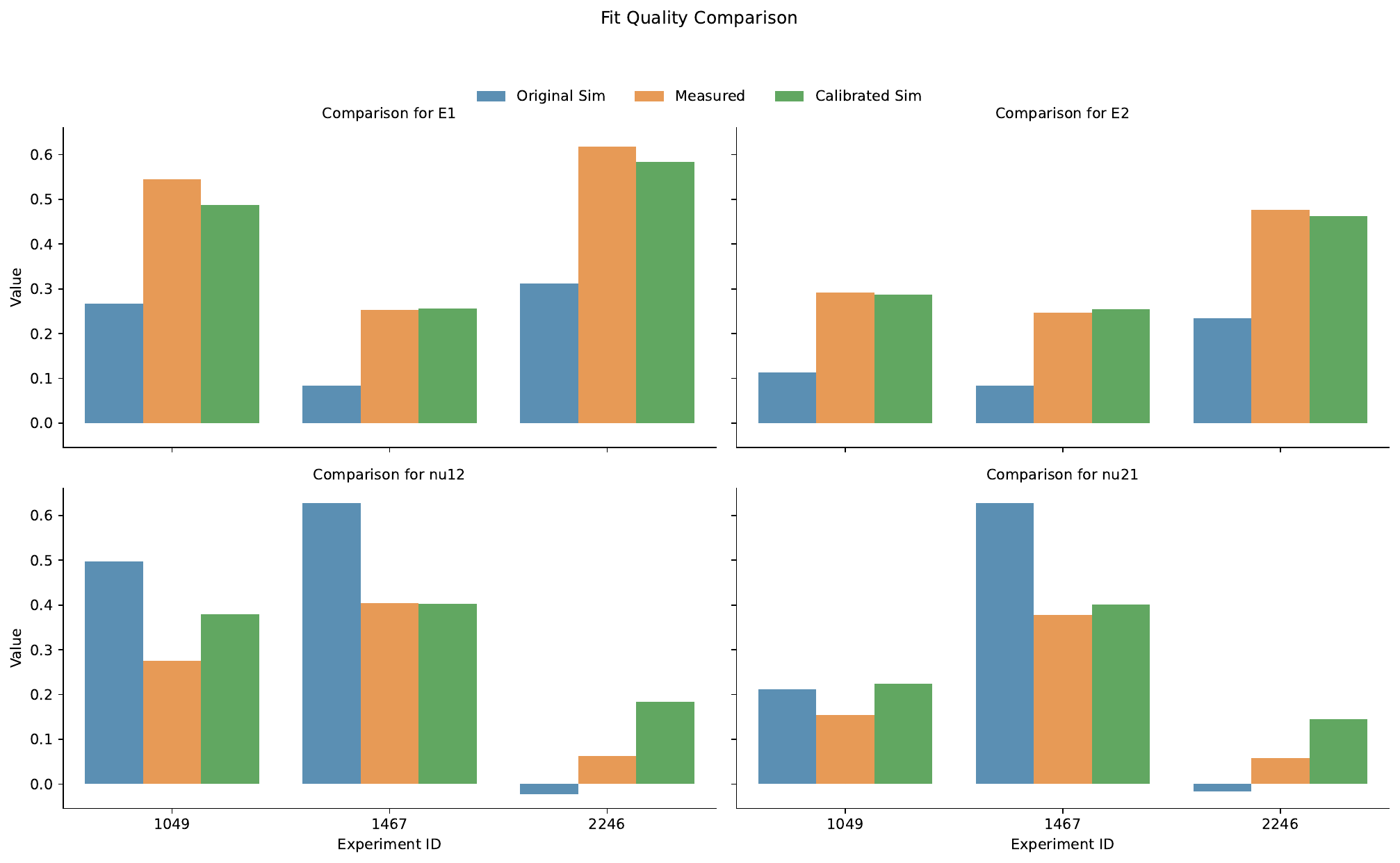}
    \caption{Comparison of simulated, measured, and calibrated material parameters for $E_1$, $E_2$, $\nu_{12}$, and $\nu_{21}$. The orange and green bars are the values shown in Table 1 in the main text.
    }
    \label{fig:prop_calibration_comparison}
\end{figure}

\autoref{fig:prop_calibration_comparison} quantitatively demonstrates the impact of the calibration by comparing the measured effective properties against both the original ($\chi=0.5$) and calibrated ($\chi=0.15$) simulation results for $E_1$, $E_2$, $\nu_{12}$, and $\nu_{21}$. The figure shows a clear improvement in the agreement between simulated and measured values for these properties after applying the calibrated $\chi$.

Consequently, all simulated properties reported in the main text (main text Table 1, Figure 4) were generated using the calibrated projection dilation factor of $\chi=0.15$, while other TO parameters ($E_{max}$, $E_{min}$, $\nu$, $\beta$) were kept at their initially assumed values.

\begin{table}[h]
    \centering
    \caption{Original outputs of the IH process (column 1) and the post-calibrated topologies (column 2) for the examples from Figure 2 and as described in the main text section 6.5}
    \label{tab:dilation}
    \begin{tabular}{cc}
        \toprule
        Original Topology $\chi=0.5$ & Dilated Topology $\chi=0.15$ \\
        \midrule
        \includegraphics[width=0.3\textwidth]{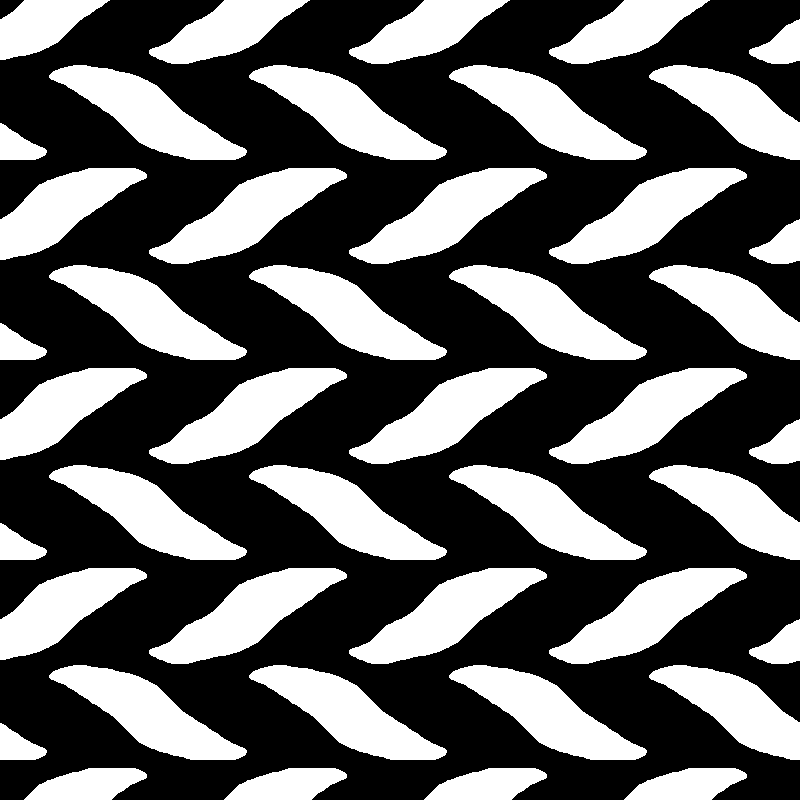} &
            \includegraphics[width=0.3\textwidth]{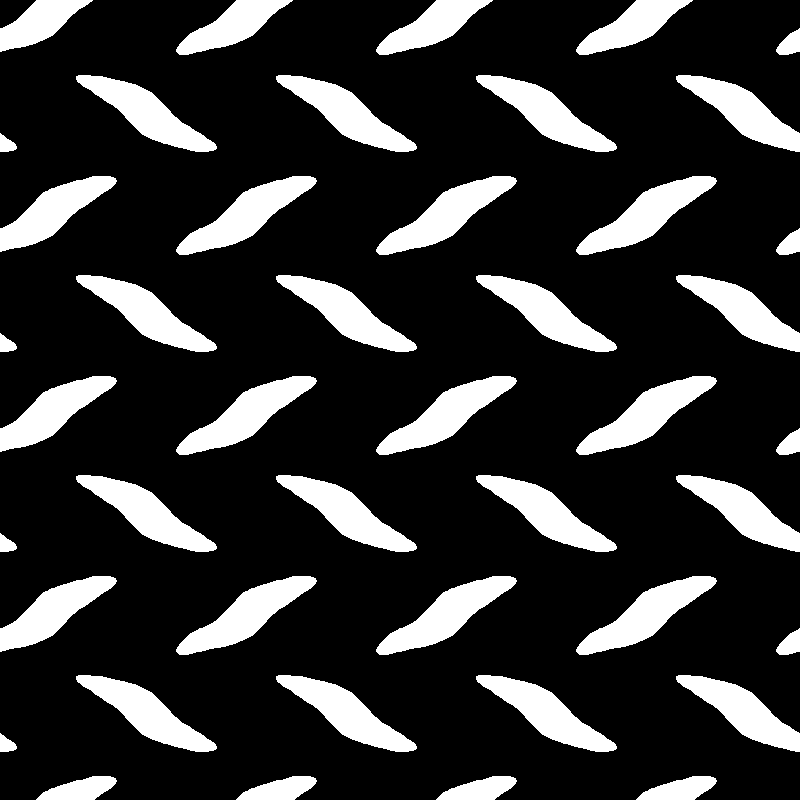} \\
        \includegraphics[width=0.3\textwidth]{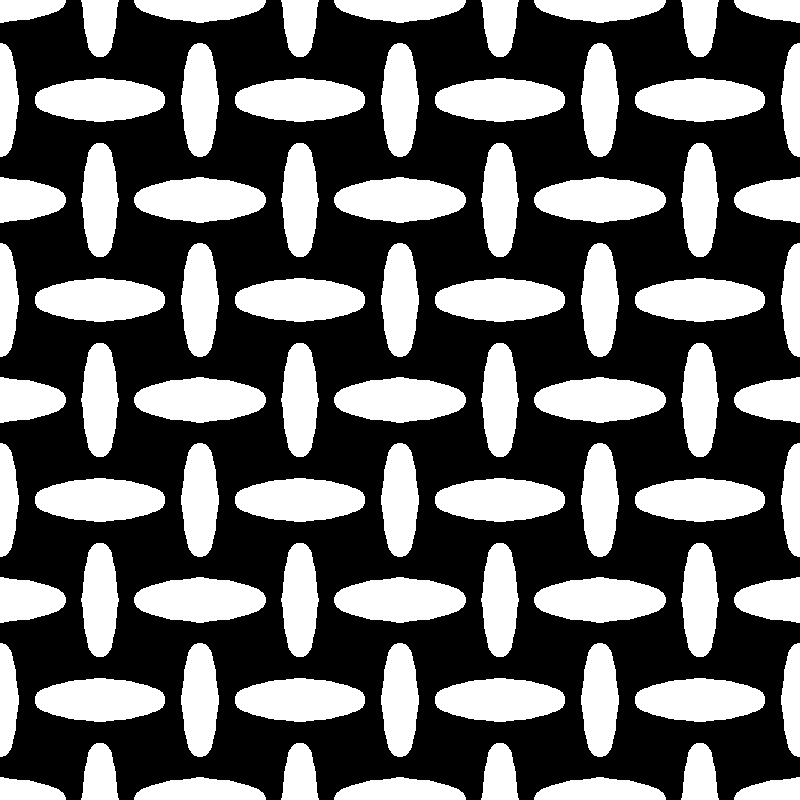} &
            \includegraphics[width=0.3\textwidth]{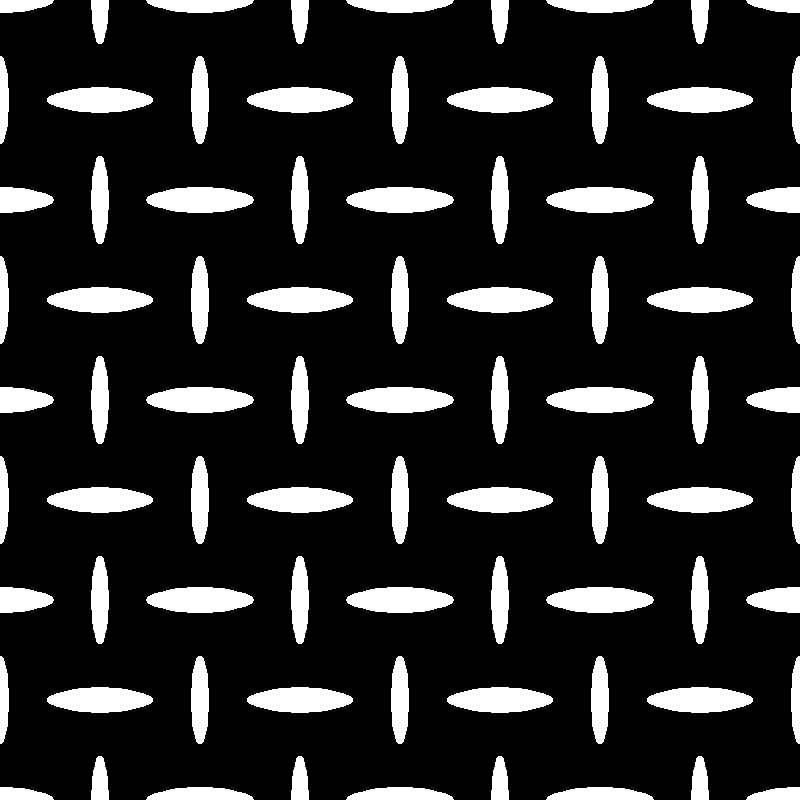} \\
        \includegraphics[width=0.3\textwidth]{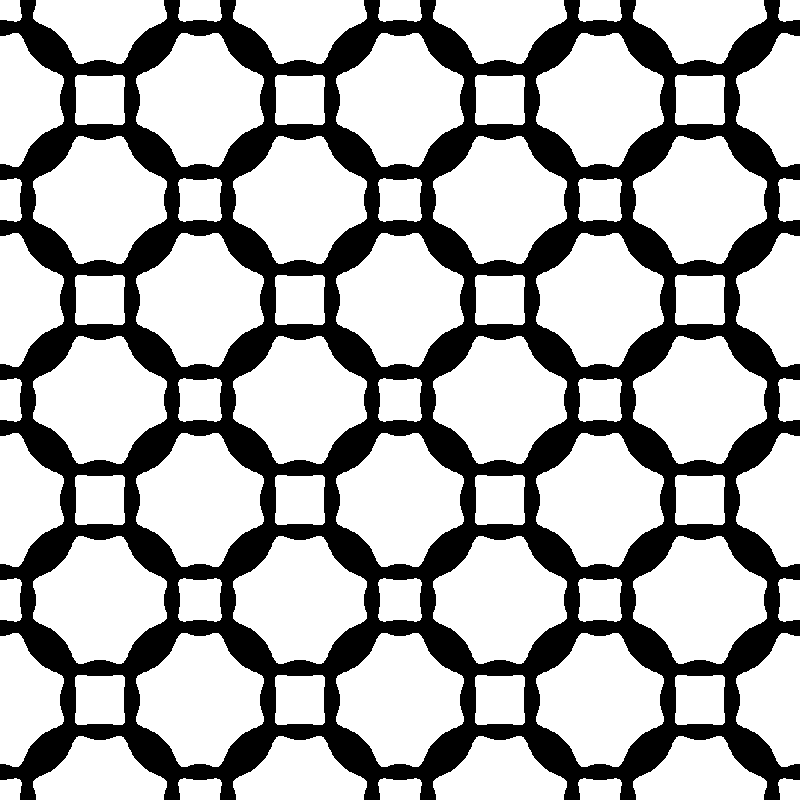} &
            \includegraphics[width=0.3\textwidth]{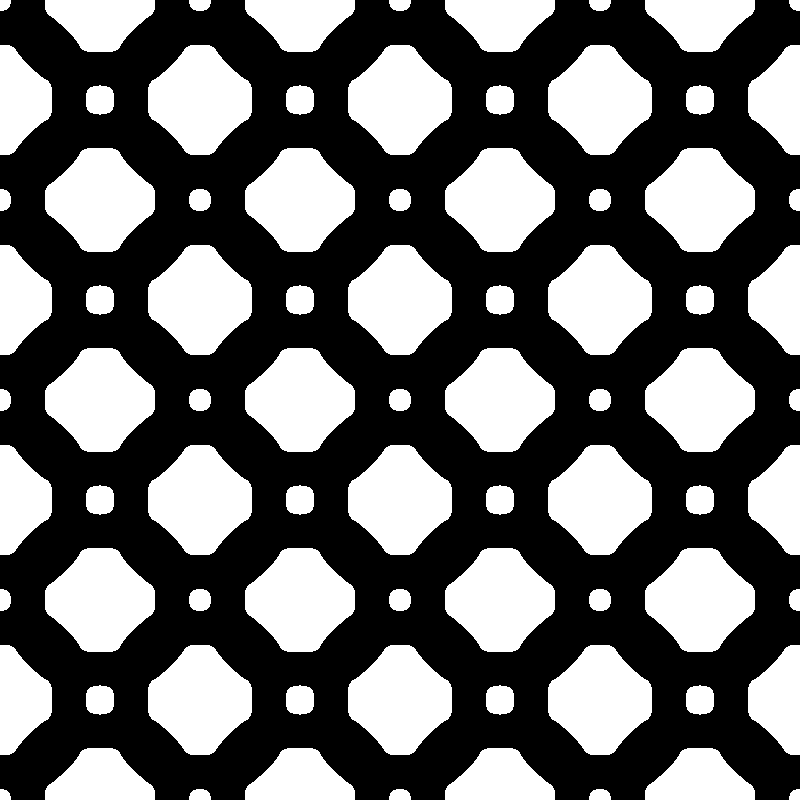} \\
        \includegraphics[width=0.3\textwidth]{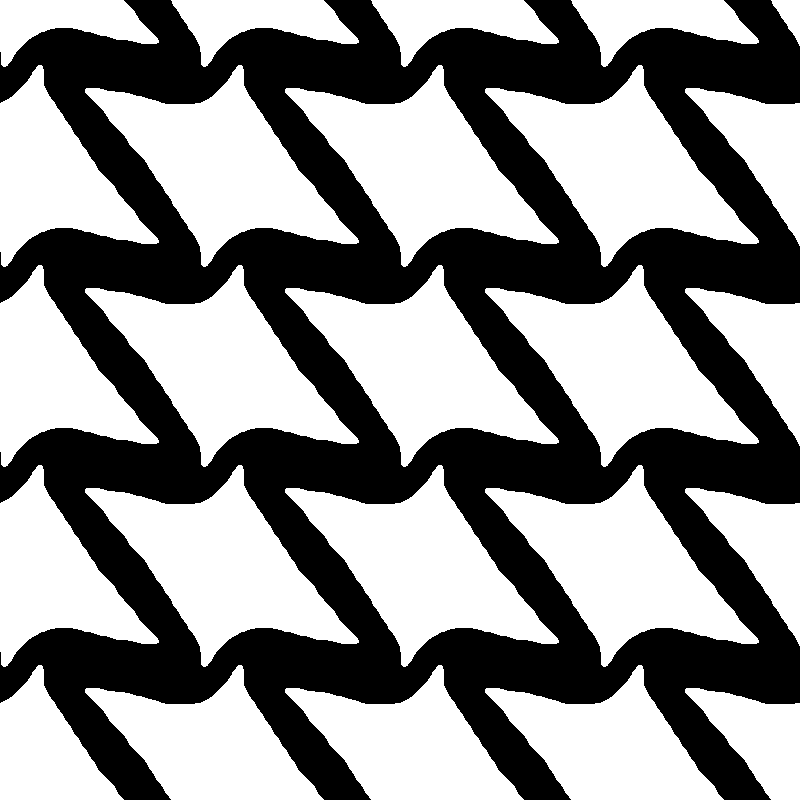} &
            \includegraphics[width=0.3\textwidth]{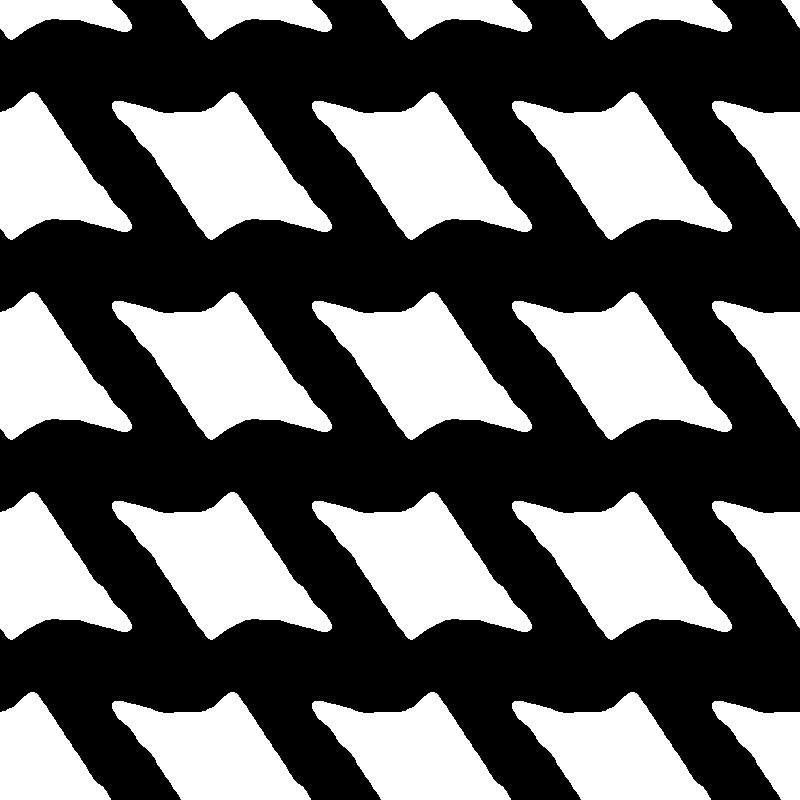}
\end{tabular}
\end{table}

\bibliography{paperpile}